\documentstyle[preprint,aps,eqsecnum]{revtex}
\tightenlines
\begin{document}

\thispagestyle{empty}
\setcounter{page}{1}

\title{On the Thermal Gauge Boson Masses of the 
Electroweak Theory in the Broken Phase}

\author{Cristina Manuel}

\address{Dpt. Estructura i Constituents de la Mat\`{e}ria\\
Facultat de F\'{\i}sica,
Universitat de Barcelona \\
Diagonal 647,
08028 Barcelona (SPAIN)} 

\maketitle

\thispagestyle{empty}
\setcounter{page}{0}

\begin{abstract}
$\!\!$Thermal effects in the  broken phase of  the electroweak
theory are  studied in the strongly interacting Higgs boson limit.
In that limit and at tree level
the bosonic sector of the theory is a gauged non-linear sigma model.
The associated  one-loop
thermal effective action for soft fields is then computed
by using  the background
field method together with the Stueckelberg formalism. 
This effective action describes thermal corrections to the masses of the gauge 
bosons $W$, $Z$ and the photon. It is the proper
generalization of the hard thermal effective action of a Yang-Mills
theory when there is a Higgs mechanism for a heavy Higgs particle.
\end{abstract}

\vfill

\hfill\break
\hbox to \hsize{ECM-UB-PF-98/01} 
\hbox to \hsize{January/1998}
\vskip-12pt
\eject

\baselineskip=15pt
\pagestyle{plain}

\section{INTRODUCTION}
\label{Intro}

This article is devoted to  study  thermal effects in the broken
phase of the electroweak model in the strongly interacting  Higgs boson limit. 
In that limit the Higgs field becomes
heavy and it can be integrated out. At tree level the corresponding 
low energy effective theory is a gauged non-linear sigma model
\cite{Appel}. 

The electroweak theory is supposed to undergo a phase transition 
at  high temperature $T$ \cite{phaseT}. The fact that the model has
a completely different behavior at low or high temperatures has important 
phenomenological consequences,  which will not be discussed
here. Instead, I will compute one-loop thermal effects
at $T$ far below the electroweak phase transition, taking profit of
some results already obtained in the contexts of QCD and ungauged non-linear
sigma models. Let us first comment about those two different field theories.

In the last few years much progress has been made in understanding
the leading thermal effects in non-Abelian gauge theories, such as QCD 
\cite{LeBellac}.
Even if at $T=0$ the non-Abelian gauge fields are massless, at finite
$T$ there is a thermal mass, the so-called Debye mass, for non-Abelian
electric fields. The Debye mass can be computed at one-loop order in
perturbation theory. To take into account
the thermal effects of Debye screening properly a resummation of
an infinite number of Feynman diagrams, the hard thermal loops (HTL's),
is required \cite{BP}, \cite{cHTL}, \cite{gaugeinv}.
Hard thermal loops are thermal amplitudes which arise in
a gauge theory when the external loop
momenta is {\it soft}, while the internal one is {\it hard}.
{\it Soft} denotes a scale $\sim gT$, while {\it hard} refers to one $\sim T$,
where $g \ll 1$ is the gauge coupling constant.
The effective action generating HTL's  is a mass term for the
chromoelectric fields, and it has been constructed 
just by solving a gauge invariance condition imposed on it \cite{gaugeinv}.
This mass term is, however, non-local.

When a global $SU(N)_L \times SU(N)_R$ symmetry is spontaneously broken to
$SU(N)_{L+R}$ the physics of the corresponding Goldstone bosons at lowest order
in their momenta is described by the (ungauged) non-linear sigma model. Thermal effects
in this model  have already been considered in the literature
\cite{fT}, \cite{Gavela}, \cite{PT}, \cite{cm1}.
Surprisingly, the same one-loop thermal amplitudes as before, the
HTL's, appear in the framework of the non-linear sigma model
\cite{PT}, \cite{cm1}. This can be
understood in terms of symmetry arguments.
In this case HTL's give account of thermal scattering among the Goldstone bosons.

I will use  the knowledge of the thermal effects of the two above mentioned
models for the low energy effective theory of the electroweak model.
The one-loop thermal effective action for soft gauge fields in this  effective
model will be computed. For the gauge fields to be soft,
one has to require that the masses of the gauge bosons be also soft, thus
$\ll T$. The temperature  range of validity of the results is then
$M_W,M_Z \ll T \ll \sqrt{12} v$, where $M_W$ and $M_Z$ are the masses of the gauge 
bosons $W$ and $Z$, respectively, and $v$ is the vacuum expectation value of the
Higgs doublet at $T=0$. The upper limit is so because in the $N=2$ non-linear sigma model
thermal corrections can be written as an expansion in the dimensionless parameter $(T^2/12 v^2)$. 

The computation will be done using the background field method\cite{BFM}
together with the Stueckelberg formalism \cite{Stu}, \cite{Ditt}.
The thermal effective action computed here turns
out to be, in the unitary gauge,  a combination of  the
one-loop thermal effective actions which appear,
on one hand, in a pure Yang-Mills theory, and on the other hand, in
a non-linear sigma model in the presence  of external currents. The result is
thus written in terms of the same non-local  thermal effective actions which appear
in the previous mentioned theories. Furthermore, 
it gives a description of the thermal corrections to the masses of 
the gauge bosons. Therefore,  it can be considered as the proper 
generalization of the hard thermal loop effective action of a non-Abelian
gauge field theory when there is a Higgs mechanism, and for a heavy Higgs
particle.

This paper is structured as follows. In Sect.~\ref{revelmd} the bosonic
sector of the electroweak model is reviewed. The corresponding
effective theory in the strongly interacting Higgs boson limit is a gauged non-linear
sigma model. In Sect.~\ref{BFMSF} the background field method and the 
Stueckelberg formalism applied to that model are presented. In
Sect.~\ref{OLTEA} the one-loop thermal effective action is obtained. From that
action the thermal corrections to the masses of the gauge bosons are
extracted in Sect.~\ref{TMSGF}. The last Section is devoted to the conclusions.

\section{Bosonic Sector of the Electroweak Theory}
\label{revelmd}

In order to set up the notation and conventions used in this article
the bosonic sector of the electroweak theory is briefly reviewed in
this Section \cite{Appel}. The classical Lagrangian of the
bosonic sector of the electroweak theory $SU(2)_W \times U(1)_Y$ is written 
in terms of the weak isospin
gauge field $W_\mu$, the weak hypercharge gauge field $B_\mu$, and the 
complex scalar doublet $\Phi$. Using the standard linear representation
for the Higgs fields, the classical Lagrangian reads
\begin{eqnarray}
\label{2.1}
{\cal L}_{cl}  & = &  {\cal L}_{gauge} + {\cal L}_{Higgs} \\
& = & - \frac 12 \,{\rm Tr} (W_{\mu \nu} W^{\mu \nu} ) - 
\frac 14 \, B_{\mu \nu} B^{\mu \nu} +  D_\mu \Phi  D^\mu \Phi^\dagger
+ \mu^2  \Phi \Phi^\dagger - \lambda  (\Phi \Phi^\dagger)^2 \ , \nonumber
\end{eqnarray}
where the $SU(2)_W$ and $U(1)_Y$ field strength are
\begin{equation}
\label{2.2}
W_{\mu \nu} = \partial_\mu W_\nu - \partial_\nu W_\mu + i g \,[W_\mu , W_\mu] \ ,
\qquad B_{\mu \nu} = \partial_\mu B_\nu - \partial_\nu B_\mu \ ,
\end{equation}
with $W_\mu = \frac 12 {\vec W}_\mu  \cdot \vec{\tau}$, and $\vec{\tau} =
(\tau^1,\tau^2,\tau^3)$ 
are  the Pauli matrices.
The covariant derivative acting on the Higgs doublet is
\begin{equation}
\label{2.3}
D_\mu \Phi = \partial_\mu \Phi + i g \, W_\mu \Phi + i \frac {g'}{2} B_\mu \Phi \ .
\end{equation}
The complex $SU(2)_W$ doublet is represented linearly as
\begin{equation}
\label{2.4}
\Phi (x) = \pmatrix{ \varphi^+ (x)\cr  \varphi^0(x)}  \ .
\end{equation}
To study the strongly interacting effects of the electroweak model it 
is more convenient to use a different representation for the Higgs
sector. After defining the matrix \cite{Appel}
\begin{equation}
\label{2.5}
M (x) = \sqrt{2} \pmatrix{\varphi^{0 \dagger}(x) & \varphi^+ (x)\cr
- \varphi^- (x) &  \varphi^0(x)} = \sqrt{2} \left( {\tilde \Phi} \,  \Phi
\right) \ ,
\end{equation}
where ${\tilde \Phi} \equiv i \tau^2 \Phi^*$,  the Lagrangian (\ref{2.1}) 
becomes
\begin{eqnarray}
\label{2.6}
{\cal L}_{cl} & = &- \frac 12 \,{\rm Tr} (W_{\mu \nu} W^{\mu \nu} ) - 
\frac 14 \,B_{\mu \nu} B^{\mu \nu} + \frac 14 \,{\rm Tr} (D_\mu M  D^\mu M ^\dagger)
\\
&  - & \frac{\lambda}{4} 
\left(\frac 12 \, {\rm Tr} (M M^\dagger) - \frac{\mu^2}{\lambda} \right)^2 \ .
\nonumber
\end{eqnarray}
The covariant derivative acting on the matrix $M$ is
\begin{equation}
\label{der}
D_\mu M = \partial_\mu M + i g \, W_\mu M - i \frac {g'}{2}\, M \,B_\mu \tau^3 \ .
\end{equation}

While the fields $\Phi$ and ${\tilde \Phi}$ transform exactly in the same
way under the $SU(2)_W$ symmetry, they transform oppositely under 
the $U(1)_Y$ symmetry. This is reflected in the explicit presence
of the $\tau^3$ matrix in the covariant derivative (\ref{der}).
Let us also recall that the ungauged Higgs Lagrangian
has a global $SU(2)_L \times SU(2)_R$ symmetry, the so-called
custodial symmetry. The custodial symmetry is explicitly broken once
the gauge interactions are turned on.

The matrix $M$ can be written in terms of the physical Higgs field $H$
and the unphysical Goldstone bosons $\phi^a$ as
\begin{equation}
\label{2.8}
M (x) = \left(v + H(x) \right) \Sigma (x) \ , \qquad 
\Sigma (x) = \exp{(i \frac{{\vec \phi} \cdot {\vec \tau}}{v})}
\end{equation}
where $v= \sqrt{\mu^2 /\lambda}$ is the vacuum expectation value.
Using the non linear representation for the Higgs sector 
one then may write the Higgs Lagrangian as
\begin{equation}
\label{2.9}
{\cal L}_{Higgs} = \frac 12 \partial_\mu H  \partial^\mu H + 
\frac {(v+ H)^2}{4} {\rm Tr} \, (D_\mu \Sigma \,D^\mu \Sigma ^\dagger)
- \frac{ \lambda}{4} (v+ H)^4 + \frac{\mu^2}{2} (v+ H)^2 \ .
\end{equation}

The linear representation of the Higgs sector
and the non-linear one are physically
equivalent, and give the same physical answers. However,
the last one is more suited to study the model in the strongly
interacting limit  $\lambda \rightarrow \infty$. In that 
limit the Higgs mass, $M_H = \sqrt{2 \lambda v^2}$, becomes large, and
the Higgs field can be integrated out. Then
the effective Lagrangian of the electroweak theory reduces 
at tree level to
\begin{equation}
\label{2.10}
{\cal L}_{eff}  = - \frac 12 \,{\rm Tr} (W_{\mu \nu} W^{\mu \nu} ) - 
\frac 14\, B_{\mu \nu} B^{\mu \nu} +
 \frac {v^2}{4} {\rm Tr} \, (D_\mu \Sigma\, D^\mu \Sigma ^\dagger)\ .
\end{equation}
That is,  the low energy effective theory for the
bosonic sector of the electroweak model
is a gauged non-linear sigma model. This effective theory is non-renormalizable.
The non-renormalizability of the effective theory indicates the sensibility
of the model to an ultraviolet cutoff, which here is interpreted as $M_H$.
Beyond tree level, loop corrections associated with virtual Higgs-boson
exchange lead to additional effective interactions. Those will not be
considered here, anyway, since the study in this article will be restricted
to thermal one-loop effects.

In the unitary gauge, that is, in the gauge where
all the unphysical Goldstone bosons are eaten by the gauge fields
(i.e. where $\Sigma =1$), one can read off
the masses of the physical gauge fields  from Eq.  (\ref{2.10}).
The fields $W_\mu ^+$, $Z_\mu$ and $A_\mu$
are defined as
\begin{eqnarray}
\label{2.11}
W_\mu ^{\mp} & = & \frac {1}{\sqrt{2}} \left(W_\mu ^1 \pm i W_\mu ^2 \right) \ , \\
\label{2.12}
Z_\mu & = & \cos{\theta_W} W_\mu ^3 - \sin{\theta_W} B_\mu \ , \\
A_\mu & = & \sin{\theta_W} W_\mu ^3 + \cos{\theta_W} B_\mu \ ,
\label{2.13}
\end{eqnarray}
where $\theta_W$ is the Weinberg angle, $\tan{\theta_W}= g'/g$. The masses of those
fields are
\begin{equation}
\label{2.14}
M_W = \frac{v g}{2} \ , \qquad M_Z = \frac{v}{2} \sqrt{g^2 + g'^2} \ , 
\end{equation}
while the mass of the photon is $ M_\gamma = 0$.
At tree level the masses of the gauge fields obey the relation
\begin{equation}
\label{2.15}
\rho \equiv (M_W / M_Z\cos{\theta_W})^2 =1
\end{equation}
 as a consequence of the custodial
symmetry.

Let us finally recall that the electric charge is defined as
\begin{equation}
\label{2.16}
e = \frac{g g'}{\sqrt{g^2 + g'^2}} \ .
\end{equation}

\section{The Background Field Method and the Stueckelberg Formalism}
\label{BFMSF}

In this Section the background field method (BFM) \cite{BFM} and the Stueckelberg
formalism \cite{Stu} for the gauged non-linear sigma model (\ref{2.10}) are described. 
Those formalisms have already been used in the context of the electroweak
model \cite{Ditt}.

According to the BFM one has to split all fields into background 
and quantum pieces. The background fields are solutions of the classical
equations of motion.
In the present case one defines an additive splitting for the gauge fields,
thus
\begin{equation}
\label{3.1}
W_\mu  (x)  = {\bar W_\mu}  (x) + w_\mu (x)  \ , \qquad B_\mu  (x) = {\bar B_\mu} 
(x)  + b_\mu  (x)  \ .
\end{equation}
The background gauge fields have been represented by capital  letters with a bar,
while the quantum ones are denoted by lower case letters. 
The Goldstone fields are split multiplicatively, thus
\begin{equation}
\label{3.2}
\Sigma  (x) = \xi (x)  h (x)  \xi (x) \ , \qquad  {\bar \Sigma} = \xi (x)  \xi  (x) \ ,
\end{equation}
where $\xi$ is a unitary matrix, $\xi \xi^\dagger = 1$.
The background field ${\bar \Sigma}$ and
the quantum field $h$ are written in terms of background and quantum
Goldstone fields, ${\bar \phi}$ and $\phi$,
 respectively, as ${\bar \Sigma}=\exp{(i {\bar \phi^a} \tau^a/v)}$ and
$h = \exp{(i \phi^a \tau^a/v)}$.

After the splitting of fields is done, the  Lagrangian (\ref{2.10}) is 
separately invariant under background and quantum gauge transformations.
Let us see what the behavior of the fields is under those two different
kinds of gauge transformations.

Under background gauge transformations the background field ${\bar \Sigma}$
transform as follows
\begin{equation}
\label{3.3}
{\bar \Sigma}' (x) = U_W (x) \Sigma (x) U_Y (x) \ ,
\end{equation}
where 
\begin{equation}
\label{3.4}
U_W (x) = \exp{(ig \frac{{\vec \theta_W} \cdot {\vec \tau}}{2})} \ , \qquad 
U_Y (x) = \exp{(ig' \frac{\theta_Y  \tau^3}{2})} \ .
\end{equation}
The field $\xi$ behaves under the same transformation as
\begin{equation}
\label{3.5}
 \xi' (x) = U_W (x) \xi (x) V^\dagger (x)=  V(x) \xi (x) U_Y (x) \ ,
\end{equation}
and $V(x)$ is a  unitary matrix which depends on $U_W$, $U_L$ and $\xi$.
The background gauge fields  transform as
\begin{eqnarray}
\label{3.6}
{\bar W}'_\mu (x)& = & U_W (x) {\bar W}_\mu (x)U_W ^\dagger (x) - \frac ig \,
U_W (x) \partial_\mu U_W ^\dagger (x) \ , 
\\
\label{3.7}
{\bar B}'_\mu (x) & = &{\bar B}_\mu (x)+ \partial_\mu \theta_Y (x) \ ,
\end{eqnarray}
while under the same transformation, the quantum gauge fields behave as
\begin{equation}
\label{3.8}
w'_\mu (x) =  U_W (x) w_\mu (x) U_W ^\dagger (x) \ , \qquad b'_\mu (x) =
 b_\mu (x) \ ,  \qquad h'(x) = V (x) h(x) V^\dagger (x) \ .
\end{equation}

Under infinitesimal quantum gauge transformations the
background   fields remain invariant
\begin{equation}
\label{3.9}
\delta {\bar W}_\mu (x) = 0 \ , \qquad \delta {\bar B}_\mu (x)= 0 \ ,  \qquad
\delta {\bar \Sigma} (x) = 0 \ ,
\end{equation} 
while the quantum gauge fields behave as
\begin{equation}
\label{3.10}
\delta w_\mu ^a = - \left(\partial_\mu \alpha_w^a - g \epsilon^{abc} ({\bar
W}_\mu ^b + w_\mu ^b) \alpha_w ^c \right) \ , \qquad \delta b_\mu = \partial_\mu 
\alpha_y \ .
\end{equation}
The quantum Goldstone field transform as
\begin{equation}
\label{3.11}
h'(x) = V_w (x) h(x) V_y (x) \ , 
\end{equation} 
where 
\begin{equation}
V_w (x) = \xi^\dagger (x) \exp{(ig \frac{{\vec \alpha_w} \cdot {\vec \tau}} {2})} 
\xi (x)  \ , \qquad V_y (x) =
\xi(x) \exp{(\frac{\alpha_y 
\tau^3}{2} )} \xi^\dagger (x) \ .
\end{equation}

In the spirit of the BFM 
the Lagrangian ${\cal L}_{eff}$ is expanded around the classical
fields, keeping terms which are quadratic in the quantum fluctuations
\begin{equation}
\label{3.12}
{\cal L}_{eff} ( W_\mu, B_\mu, \Sigma) = {\cal L}_{eff} ^{(0)} 
({\bar  W_\mu},{\bar B}_\mu, {\bar \Sigma}) + {\cal L}_{eff} ^{(2)}
({\bar  W_\mu},{\bar B}_\mu, {\bar \Sigma}, w_\mu, b_\mu, \phi) + .... 
\end{equation}
The linear terms in the quantum fluctuations vanish if one imposes the
classical equations of motion for the background fields.
To derive the one-loop effective action one has to integrate out the
quantum fields appearing in ${\cal L}_{eff} ^{(2)}$, adding the
corresponding quantum gauge-fixing and  quantum Faddeev-Popov terms.
The functional integral can be done since the dependence on the fields is only
quadratic.  The background fields  which appear in ${\cal L}_{eff} ^{(2)}$ only
act as sources for the generation of vertex functions in the effective action.
If instead of making the  functional integral
one chooses a  diagrammatic approach to get the one-loop effective action,
then one only needs to consider  Feynman diagrams
with quantum fields running inside the loop, while the background fields
only generate  external vertices.

One of the advantages of using the BFM in a gauge field theory is that
it gives a gauge invariant effective action in the background fields
right away. To do so one only needs to fix the gauge of the quantum fields
in a way invariant under the background gauge transformations. Then the
associated Faddeev-Popov operators are also invariant under 
background gauge transformations.

In the BFM  it is possible to fix the 
background and quantum gauges independently. In our case, 
and to simplify the computations, it is convenient to choose the
unitary gauge for the background  fields. Then the background 
Goldstone fields disappear completely from the Lagrangian, 
since they are eaten by the background gauge fields to become 
massive. In order to do so it is convenient to use the 
Stueckelberg formalism \cite{Stu}. 
The Stueckelberg formalism allows one to eliminate
the background Goldstone fields from the Lagrangian, thus it is 
equivalent to  choosing the unitary background gauge. 
As  will become
clear, after performing a Stueckelberg transformation the fields
${\bar \Sigma}$ and $\xi$ are mapped to the unit matrix, and anything else
remains unaffected.
To see how this can be done, let us consider the following
term of the Lagrangian (\ref{2.10})
\begin{equation}
\label{3.13}
\frac{v^2} {4} \, {\rm Tr} (D_\mu (\xi h \xi) D^\mu (\xi h \xi)^\dagger) \ .
\end{equation}
The covariant derivative appearing in Eq. (\ref{3.13}) can be written as
\begin{equation}
\label{3.14}
D_\mu (\xi h \xi) = \xi \left( \partial_\mu h + \xi^\dagger  \nabla_\mu ^W \xi \,h
- h \,\xi  \nabla_\mu ^B \xi^\dagger \right) \xi  \ ,
\end{equation}
where 
\begin{eqnarray}
\label{3.15}
\xi^\dagger  \nabla_\mu ^W \xi & =& \xi^\dagger \left( \partial_\mu \xi + i g
({\bar W}_\mu + w_\mu ) \xi \right) \ , \\
\label{3.16}
\xi  \nabla_\mu ^B \xi^\dagger & =& \xi \left( \partial_\mu \xi^\dagger 
+ i g' ( {\bar B}_\mu + b_\mu )\frac {\tau^3}{2} \xi^\dagger \right) \ . 
\end{eqnarray}

Then if one performs the Stueckelberg transformation
\begin{eqnarray}
\label{3.17}
&{\bar W}'_\mu  =  \xi^\dagger  {\bar W}_\mu \xi - \frac ig \,\xi^\dagger
\partial_\mu \xi \ , \,&
{\bar B}'_\mu \frac {\tau^3}{2}  = \xi ({\bar B}_\mu \frac {\tau^3}{2} )\xi^\dagger
- \frac {i}{g'}\, \xi \partial_\mu \xi^\dagger \ , \\
\label{3.18}
& w'_\mu  = \xi^\dagger  w_\mu \xi \ , &
b'_\mu  \frac {\tau^3}{2}  =  \xi (b_\mu  \frac {\tau^3}{2}) \xi^\dagger \ ,
\end{eqnarray}
and one writes the Lagrangian in terms of the primed fields, all the
background Goldstone fields $\xi$ disappear completely!
 It is easy to check that after the transformation (\ref{3.17}-\ref{3.18})
is done,  Eq. (\ref{3.13}) becomes
\begin{equation}
\label{3.19}
\frac{v^2} {4} \, {\rm Tr} (D'_\mu h\,  D'^\mu  h^\dagger) \ .
\end{equation}

It has also to be pointed out that the Jacobian of the change of variables
(\ref{3.18}) is one at one-loop order. 

The Stueckelberg transformation simplifies drastically the one-loop computations.
Once the computation is finished, the Stueckelberg transformation 
has to be inverted to recover the presence of the background Goldstone
bosons in the final one-loop effective action.

In order to simplify the notation from now on  I will omit the primes in the fields,
keeping in mind that  the transformation has
to be inverted at the end of the computation.

To integrate out the quantum fields a quantum gauge fixing condition 
invariant under the background gauge transformation has to be given.
In the unitary background gauge
the gauge fixing condition for the quantum fields is chosen as
\begin{equation}
\label{3.20}
{\cal L}_{gf}^{(2)} = - \frac{1}{a_w} \, {\rm Tr} \left({\bar D}_W ^\mu w_\mu
- \frac 14 a_w g v \phi \right)^2 -   \frac{1}{2 a_b} 
\left(\partial^\mu b_\mu + \frac 12 a_b g' v \phi_3 \right)^2 \ ,
\end{equation}
where $a_w$ and $a_b$ are the gauge fixing parameters.
These gauge fixing terms are chosen such as to cancel the
unwanted pieces $\partial^\mu b_\mu \phi_3$ and 
${\rm Tr} (\partial^\mu w_\mu \phi)$ in  ${\cal L}_{eff}^{(2)}$. The form
of the gauge fixing term in an arbitrary background gauge can be 
obtained by inverting the Stueckelberg transformation.

The Faddeev-Popov terms associated to  the gauge fixing (\ref{3.20}) are 
computed as usual. Finally, the complete one-loop quantum Lagrangian 
reads in Minkowski space
\begin{eqnarray} 
\label{3.21}
{\cal L}^{(2)}_{eff} + {\cal L}_{gf}^{(2)} +{\cal L}_{FP}^{(2)}
 & = & {\rm Tr}\left(  w_\mu (g^{\mu \nu} {\bar D}^2 _W + 
\frac{1 - a_w}{a_w} {\bar D}_W^\mu {\bar D}_W^\nu + 2 i g {\bar W}^{\mu \nu} )
 w_\nu 
\right)   \\
& + & \frac 12 b_\mu \left( g^{\mu \nu} \partial^2 + \frac{1 - a_b}{a_b}
\partial^\mu \partial^\nu \right) b_\nu  \nonumber \\
&  + & M^2_W \, {\rm Tr}  (  w_\mu   w^\mu) + \frac {M^2_B}{2}\, b_\mu b^\mu - g g'
v^2 w_\mu ^3
b^\mu  \nonumber \\
&+ & \frac 14 \, {\rm Tr} ({\bar d}_\mu \phi)^2 - \frac 14 \, {\rm Tr} [{\bar \Delta}_\mu, \phi]^2 
- \frac{a_w M^2_W}{4} {\rm Tr} \phi^2 - \frac{a_b M^2_B}{2} \phi^2_3 \nonumber \\
&+ &  2 v\, {\rm Tr} \left( (g w_\mu - g' b_\mu \frac{\tau^3}{2}) {\bar \Gamma}^\mu \phi \right) \nonumber \\
& - & \eta^{\dagger}_a \left(\delta^{a b}  {\bar D}^2 _W + \delta^{a b} a_w M^2_W \right) \eta_b
 \nonumber 
\end{eqnarray}
where $M^2_B = g'^2 v^2/4$ and
\begin{eqnarray}
\label{3.22}
{\bar D}_W^\mu & = & \partial^\mu + ig [ {\bar W}^\mu, \, ] \ , \\
\label{3.23}
{\bar d}_\mu \phi & = & \partial_\mu \phi + [ {\bar \Gamma}_\mu, \phi] \ , \\
\label{3.24}
{\bar \Gamma}_\mu & = & \frac i2 \left( g {\bar W}_\mu + g' {\bar B}_\mu
\frac{\tau^3}{2} \right) \ , \\
\label{3.25}
{\bar \Delta}_\mu & = & \frac i2 \left( g {\bar W}_\mu - g' {\bar B}_\mu
\frac{\tau^3}{2} \right) \ . 
\end{eqnarray}

The ghost fields $\eta_a$ are 
associated to the $w_\mu ^a$ quantum
fields. Since the ghost associated to the $b_\mu$ field does not couple
to any background external field, it has been omitted in Eq. (\ref{3.21}).

The one-loop Lagrangian (\ref{3.21}) is written in the unitary background
gauge. It can be obtained in an arbitrary background gauge  by
inverting the Stueckelberg transformation. However, it is much simpler
to integrate out the quantum fields first, and invert the transformation
afterwards to obtain the one-loop effective action in a general 
background gauge. That is what it will be done in the following Section.

\section{One-Loop Thermal Effective Action for Soft Modes}
\label{OLTEA}

\subsection{Unitary Background Gauge}
In this Section one-loop thermal effects are derived using the BFM 
Lagrangian (\ref{3.21}) in an arbitrary quantum
 gauge $a_w , a_b \neq 0, \infty$. In those gauges the quantum
fields $w_\mu^a$, $b_\mu$, $\phi^a$ and also the ghosts are massive.
Only thermal effects will be studied, and the 
$T=0$ results will not be taken into account. I will actually take profit
of results already derived in the literature to get the thermal effective
action. Those results were obtained using the imaginary time formalism.

The analysis will be restricted to soft background fields ${\bar W}_\mu$,
${\bar Z}_\mu$ and ${\bar A}_\mu$. Those fields can only be soft if their
respective masses are also soft, thus $M_W, M_Z \ll T$. This, in turn,
implies that $g^2, g'^2 \ll (T^2 / v^2)$.

For soft background fields the leading thermal corrections arise when
the internal quantum fields are hard \cite{BP}, that is, of energy $\sim T$. If one
neglects corrections of order $M_W /T$ and $M_Z /T$ in the final answers,
then it is possible to neglect those masses for the quantum fields. 
In other words, for hard quantum fields the
terms $\partial^2$ of the Lagrangian are of the order $T^2$, which are dominant as compared
to the terms $M^2_{W,B}$, which therefore will be neglected.

The computation simplifies once the masses of the quantum fields are neglected. 
One encounters here the same one-loop thermal amplitudes,
the HTL's, which appear in the BFM of Yang-Mills theories
\cite{BP}, as well as in the non-linear sigma model in the presence
of external background sources \cite{cm1}. There are also new types of vertices 
in (\ref{3.21}), which do not appear in the BFM studies of the  previous mentioned theories:
those which couple quantum gauge fields and quantum Goldstone bosons.
However, a power counting analysis shows that the one-loop 
thermal corrections generated by those vertices
are subleading as compared to the HTL's, and therefore they will be
neglected.

The one-loop thermal effective action for soft background gauge fields
is then a combination of the one which appears in a Yang-Mills theory
and the one in the non-linear sigma model in the presence of external sources.
I refer to the literature to see how those effective actions are computed \cite{BP},
\cite{gaugeinv}, \cite{cm1}.
By translating those results to our case one  finds the following
one-loop thermal effective action

\begin{eqnarray}
\label{4.1}
S_{eff} + \delta S_{eff,T} & = & \int d^4 x \left\{
- \frac 12 \,{\rm Tr} ({\bar W}_{\mu \nu} {\bar W}^{\mu \nu} ) - 
\frac 14 \, {\bar B}_{\mu \nu} {\bar B}^{\mu \nu} + \frac {v^2(T)}{4} 
\,{\rm Tr}\left( g {\bar W}^\mu - g' {\bar B}^\mu \frac{\tau^3}{2} \right)^2 \right\}
\\
&- &\frac{T^2}{6}\int \frac{d \Omega_{\bf q}}{4 \pi} \int d^4 x \,d^4 y\,
{\rm Tr} \left({\bar \Gamma}_{\mu \lambda} (x) <x | \frac{Q^\mu Q_{\nu}}{- (Q \cdot
{\bar d})^2} |y> {\bar \Gamma}^{\nu \lambda} (y) \right) \nonumber \\
& + &  \frac{g^2 T^2}{3}\int \frac{d \Omega_{\bf q}}{4 \pi} \int d^4 x \,d^4 y\,
{\rm Tr} \left({\bar W}_{\mu \lambda} (x) <x | \frac{Q^\mu Q_\nu}{- (Q \cdot
{\bar D}_W)^2} |y> {\bar W}^{\nu \lambda} (y) \right)
\nonumber
\end{eqnarray}
where ${\bar W}_{\mu \nu}$,  ${\bar B}_{\mu \nu}$ are the field strengths
of the corresponding background gauge fields, and
\begin{eqnarray}
\label{4.2}
v(T) & =  & v \left( 1 - \frac{1}{12} \frac {T^2}{v^2} \right) \ , \\
\label{4.3}
{\bar \Gamma}_{\mu \nu} & = & \partial_\mu {\bar \Gamma}_\nu
- \partial_\nu {\bar \Gamma}_\mu +
 [{\bar \Gamma}_\mu,{\bar \Gamma}_\nu] \ , 
\end{eqnarray}
and $Q= (i, {\bf q})$ is a null vector $Q^2=0$. The angular
integral in (\ref{4.1}) is done over all directions of the three 
dimensional unit vector ${\bf q}$.

Let me remind the meaning of each term of Eq. 
(\ref{4.1}). The two first terms are
the kinetic pieces for the soft background gauge fields.
The last piece in Eq. (\ref{4.1})
is the HTL effective action for the non-Abelian gauge field ${\bar W}_\mu$,
and it is generated by considering the one-loop thermal effects of the hard
quantum gauge field $w_\mu^a$, and the quantum ghosts $\eta^a$ \cite{BP}.
The third and fourth terms in Eq. (\ref{4.1})
arise after considering the one-loop thermal effects of the hard quantum Goldstone bosons
$\phi^a$,
(see Section III of
 Ref.\cite{cm1} with the following identifications: $F^R_\mu = -g {\bar W}_\mu$,
$F^L_\mu = -g' {\bar B}_\mu \frac{\tau^3}{2}$, and $\xi=\xi^\dagger=1$.).

\vskip1cm

\subsection{Inverting the Stueckelberg transformation}

The thermal effective action  
(\ref{4.1}) is given in
the unitary background gauge. To recover the presence of the background
Goldstone bosons in the effective action the Stueckelberg transformation (\ref{3.17}) has to
be inverted. Recall that the gauge fields which appear in Eq. 
(\ref{4.1}) are the primed
fields of Eq. (\ref{3.17}).

After the inversion of the Stueckelberg transformation one gets the effective
action
\begin{eqnarray}
\label{4.4}
S_{eff} + \delta S_{eff,T} & = & \int d^4 x \left\{
- \frac 12 \,{\rm Tr} ({\bar W}_{\mu \nu} {\bar W}^{\mu \nu} ) - 
\frac 14 \, {\bar B}_{\mu \nu} {\bar B}^{\mu \nu} + \frac {v^2(T)}{4} 
\, {\rm Tr} \, ({\bar D}_\mu {\bar \Sigma} {\bar D}^\mu {\bar \Sigma} ^\dagger) \right\}
\\
&- &\frac{T^2}{6}\int \frac{d \Omega_{\bf q}}{4 \pi} \int d^4 x \,d^4 y\,
{\rm Tr}\left({\bar \Gamma}_{\mu \lambda} (x) <x | \frac{Q^\mu Q_{\nu}}{- (Q \cdot
{\bar d})^2} |y> {\bar \Gamma}^{\nu \lambda} (y) \right)  \nonumber\\
& + & \frac{g^2 T^2}{3}\int \frac{d \Omega_{\bf q}}{4 \pi} \int d^4 x \,d^4 y\,
{\rm Tr} \left({\bar W}_{\mu \lambda} (x) <x | \frac{Q^\mu Q_\nu}{- (Q \cdot
{\bar D}_W)^2} |y> {\bar W}^{\nu \lambda} (y) \right) \ .
\nonumber
\end{eqnarray}
where now the ${\bar W}^{\mu}$ field which enters in ${\bar W}^{\mu \nu}$ and 
${\bar D}_W$ is the right hand side of  Eq. (\ref{3.17}),
and  the ${\bar \Gamma}_{\mu}$  field entering in ${\bar \Gamma}_{\mu \nu}$ and in
${\bar d}$ is 
\begin{equation}
\label{4.5}
{\bar \Gamma}_{\mu} = \frac 12 \left( \xi^\dagger  {\bar \nabla}_\mu ^ W \xi +
\xi {\bar  \nabla}_\mu ^B \xi^\dagger \right) =\frac 12 \left(\xi^\dagger \left( \partial_\mu \xi + i g
{\bar W}_\mu  \xi \right)+ \xi \left( \partial_\mu \xi^\dagger 
+ i g' {\bar B}_\mu \frac {\tau^3}{2} \xi^\dagger \right) 
 \right) \ .
\end{equation}

Equation (\ref{4.4}) gives the one-loop effective action
for soft background modes written in a general background gauge
as a function of the background gauge fields and the background 
would-be Goldstone bosons.

\section{Thermal Masses for Soft Gauge Fields}
\label{TMSGF}

The one-loop thermal effective action derived in Sec.~\ref{OLTEA}
describes gauge invariant mass terms for the gauge bosons
$W_\mu^+$, $Z_\mu$ and $A_\mu$.
In order  to read off  the thermal
corrections to those masses from the effective action 
it is convenient to use the unitary gauge, where the unphysical Goldstone
bosons are not present.

Let us consider the effective action (\ref{4.1}) in the static
limit. In the static limit the non-local terms of Eq.(\ref{4.1}) become 
local.  Apart from the kinetic terms for the gauge fields the
Lagrangian reads 

\begin{equation}
\label{5.1}
\delta {\cal L}_{eff,T}^{static} =
\frac {v^2(T)}{4} 
\,{\rm Tr}\left( g {\bar W}^\mu - g' {\bar B}^\mu \frac{\tau^3}{2} \right)^2
+ \frac{T^2}{12} \,{\rm Tr}\left( g {\bar W}_0 + g' {\bar B}_0 \frac{\tau^3}{2} \right)^2
+  \frac{2 g ^2T^2}{3} {\rm Tr}({\bar W}_0)^2 \ .
\end{equation}
Let us remark that the  thermal corrections to the two first terms of
(\ref{5.1}) were obtained by integrating
out the hard would-be quantum Goldstone bosons.

If one expresses Eq. (\ref{5.1}) in terms of the physical fields ${\bar W}^+$, ${\bar Z}_\mu$
and ${\bar A}_\mu$,  one obtains
\begin{eqnarray}
\delta {\cal L}_{eff,T}^{static} & =
&\frac{g^2 v^2 (T)}{4} {\bar W}^+ _\mu {\bar W}^{- \mu} + \frac{(g^2+g'^2) v^2 (T)}{4} {\bar Z}_\mu
{\bar Z}^\mu  \\
&+&\frac{g^2 T^2}{12}  {\bar W}^+_0 {\bar W}^{- 0} +  \frac{(g^2+g'^2) T^2 }{24}
\left( (\cos^2{\theta_W} -\sin^2{\theta_W})^2 {\bar Z}^0 {\bar Z}_0 \right. \nonumber \\
&+& \left.
4 \cos^2{\theta_W} \sin^2{\theta_W}{\bar A}_0 {\bar A}^0 + 4 \cos{\theta_W} \sin{\theta_W}
(\cos^2{\theta_W}-\sin^2{\theta_W}) {\bar Z}^0 {\bar A}_0  \right)  \nonumber \\
&+&\frac{2 g^2 T^2}{3}  {\bar W}^+_0 {\bar W}^{- 0} +   \frac{g^2 T^2 }{3}
\left( \cos^2{\theta_W} {\bar Z}^0 {\bar Z}_0 +  \sin^2{\theta_W}{\bar A}_0 {\bar A}^0 +
2 \cos{\theta_W} \sin{\theta_W} {\bar Z}^0 {\bar A}_0  \right) \ . \nonumber
\end{eqnarray}

The longitudinal and transverse gauge modes get different thermal corrections to their masses.
The thermal masses for the transverse modes are
\begin{equation}
\label{4.x}
M^2_{W,t} (T) = \frac{g^2 v^2 (T)}{4} \ , \qquad M^2_{Z,t}(T) = \frac{(g^2 + g'^2) v^2 (T)}{4} \ , 
\qquad  M^2_{\gamma, t} (T) = 0   \ ,
\end{equation}
while  for the longitudinal ones are
\begin{eqnarray}
\label{4.xx}
M^2_{W,l} (T) & = & \frac{g^2 v^2 (T)}{4} + \frac{3 g^2 T^2}{4}
\ ,  \\
 M^2_{Z,l}(T) & = & \frac{(g^2 + g'^2) v^2 (T)}{4} + 
\frac{  (g^2+g'^2) T^2}{12} \left( \cos^2{\theta_W} -\sin^2{\theta_W}\right)^2 
+ \frac{2  g^2 T^2}{3} \cos^2{\theta_W}  \ ,
 \\
\label{5.6}
M^2_{\gamma, l} (T) & = & e^2 T^2  \ .
\end{eqnarray}
To express the  electric thermal mass of the photon in terms of the electric charge $e$, use of the relation 
(\ref{2.16}) has been made.  The above results agree with those computed in Ref.
\cite{Gavela}.

Let me point out that in this article it has been considered that $M_W, M_Z \ll T$, which
implies $g^2, g'^2 \ll T^2/v^2$. It follows that the thermal corrections coming from 
integrating out the hard quantum Goldstone bosons are dominant as compared to those arising
by integrating out the hard quantum gauge degrees of freedom.
With this in mind, one can generally claim that while the transverse modes masses
decrease with $T$,  the longitudinal ones may increase with $T$, as a consequence of the
Debye screening phenomena.
It is also interesting to point out that while the transverse masses obey the relation (\ref{2.15}),
the longitudinal ones do not, once again, as a result of the thermal Debye screening.

\section{CONCLUSIONS}

The cornerstone of the electroweak model is the spontaneous breaking of the gauge
symmetry. The would-be Goldstone bosons associated to that breaking are eaten by the gauge
fields to become massive, through the standard Higgs mechanism. If the Higgs mass is very
large,  one can remove the Higgs field from the action to finally obtain a low energy effective theory.

One of the advantages of the low energy theory is that at leading order  is universal once the
pattern of the  $SU(2)_W \times U(1)_Y$ symmetry breaking is known. The  model dependence only
arises at one-loop level, through the dependence on the ultraviolet cutoff, and the matching conditions
between  the low and high energy theories. 

In this article the one-loop computations have been done putting the natural ultraviolet cutoff
$M_H \rightarrow \infty$. A finite value of $M_H$, with $T \ll M_H$, will introduce an explicit
dependence of the thermal gauge boson masses on $M_H$. These corrections to
Eqs. (\ref{4.x}-\ref{5.6}) would go as 
$\sim - T^2 \exp{(-M_H/T)}$ and  are therefore subleading.

Some general statements can be made:
\begin{itemize}
\item The would-be Goldstone bosons give a  negative thermal contribution in the transverse or
magnetic  masses.  This has  been checked at the one-loop level.

\item There is a Debye screening phenomena taking place, since the model contains charged particles.
The charged particles yield a thermal contribution which increases the value of the longitudinal or
electrical  masses with  $T$.
\end{itemize}

One should also note that if there were magnetic charges in the model, either elementary or composed 
objects, a dual Debye screening phenomena  would take place \cite{Gross}, \cite{cm2},
and  as a result the gauge transverse
masses should increase with $T$. Those effects would not be computable in perturbation
theory, though.

An important issue to be discussed  is the one of resummation. In QCD resummation of HTL's
is mandatory for soft gauge fields, as shown by using a power counting analysis \cite{BP}.
In the broken
phase of the electroweak model the same kind of arguments could be applied for the soft gauge
fields: tree graphs and the one-loop thermal amplitudes described by the effective action
(\ref{4.1}) are equally important, and resummation should be done. The implementation of the
resummation program in the electroweak model seems to be much more complicated than in QCD 
due to the presence of the would-be Goldstone bosons.  However, an advantage of
the electroweak model is that the perturbative approach should be free of infrared problems, 
contrary to what happens in QCD, due to the 
existence of thermal magnetic masses.
I hope to report in a future project  on the 
implementation of the resummation program in the electroweak theory.

\vskip 1cm

{\bf Acknowledgments:}

I am specially thankful to N. Rius for many helpful e-mail discussions in the
course of this project. I have also benefited from useful comments of 
D. Espriu and D. Litim.
This work has been supported  by funds provided by the
CICYT contract AEN95-0590,  the CIRIT contract GRQ93-1047, and the
NATO grant CRG-970055.

\end{document}